\documentstyle[11pt,newpasp,epsf,twoside]{article}
\def\edcomment#1{\iffalse\marginpar{\raggedright\sl#1\/}\else\relax\fi}
\reversemarginpar

\begin{document}
\title{The UCT CCD CV Survey}

\author{Patrick A. Woudt and Brian Warner} 

\affil{Department of Astronomy, University of Cape Town, Rondebosch 7700, South Africa}

\begin{abstract}
Some results from a high speed photometric survey of faint 
southern CVs are presented, including 7 new orbital periods.
\end{abstract}

\section{Introduction}

We are carrying out a high speed photometric survey of faint southern
Cataclysmic Variable stars (CVs) down to V = 21$^m$. The 1.0-m and 1.9-m 
telescopes at the Sutherland site of the South African Astronomical 
Observatory, in combination with the University of Cape Town (UCT) CCD camera, 
have been used to investigate the photometric behaviour of previously
unstudied faint southern Nova Remnants (NR) and Dwarf Novae (DN).
This survey -- about 35\% completed to date -- fills in a gap. Southern NRs,
are severely underobserved even though they make up the majority of the
known NRs. 

This survey aims to sample the southern CVs in order to improve the database of 
objects (in particular eclipsing systems) both for statistical purposes
and for more detailed structural studies with 8-m class telescopes.

\section{Results}

For four of the 
22 NRs that we observed with high speed photometry, we have obtained
orbital periods. This significanlty increases the number of known orbital periods for 
southern NRs. Two of these four NRs are eclipsing systems and in addition show
clear superhumps (V630 Sgr and RR Cha). We have furthermore observed 11 DN, and 
obtained a measure of the orbital period for three of them (AO Oct, XZ Eri and V359 Cen).

\begin{figure}[h]
\plotfiddle{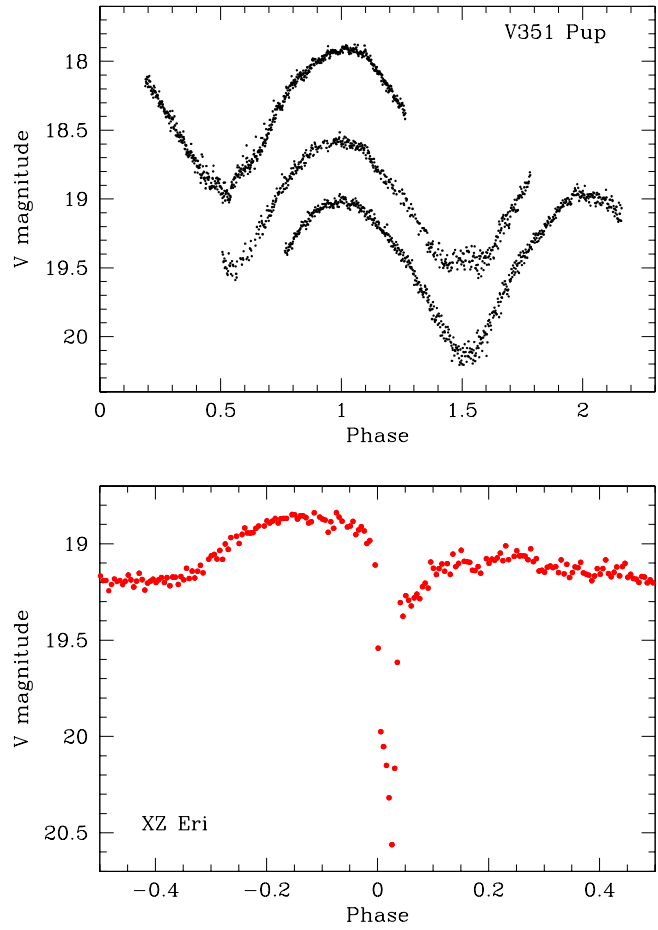}{12.0cm}{0}{60}{60}{-190}{-85}
\caption{The light curve of V351 Pup (upper panel) and XZ Eri (lower panel).}
\label{woudtf1}
\end{figure}

Fig.~1 shows two examples of light curves obtained: a NR (V351 Pup), Nova Puppis 1991,
and a DN (XZ Eri). The light curve of V351 Pup closely resembles that of the magnetic nova
V1500 Cyg (Nova Cygni 1975) at a similar point on its decline from eruption. 
XZ Eri is an eclipsing system with a very short orbital period (88.1 min), and we expect 
it to show superoutbursts and superhumps.
The orbital periods obtained so far in this survey are given in Table 1.

\begin{table}
 \centering
  \caption{Orbital periods.}
  \begin{tabular}{l c c c}
\\[2pt]
\hline \\ \vspace{1mm}
Object & Type & $<V>$ & Period \\[5pt]
\hline \hline \\[1pt] 
RS Car   & NR & 19.0 & 1.98 h  \\
V359 Cen & DN & 18.7 & 1.87 h  \\
RR Cha   & NR & 18.2 & 3.37 h  \\
XZ Eri   & DN & 19.1 & 1.47 h  \\
AO Oct   & DN & 20.5 & 1.57 h  \\
V351 Pup & NR & 18.5 & 2.84 h  \\
V630 Sgr & NR & 17.7 & 2.83 h  \\[5pt]
\hline
\end{tabular}
\end{table}

\bigskip
\acknowledgements{Our research has been supported by grants from the University of 
Cape Town.}

\end{document}